\documentclass[useAMS,usenatbib]{mn2e}

\usepackage[pdftex]{graphicx}
\usepackage[usenames,dvipsnames]{color}
\usepackage{amssymb}
\usepackage{times}
\usepackage{aas_macros}
\usepackage[normalem]{ulem}
\usepackage{marvosym}
\usepackage{url}
\usepackage[linktocpage=true,backref=true,pagebackref=false,bookmarks=true,bookmarksnumbered=False,colorlinks=true,linkcolor=NavyBlue,citecolor=NavyBlue,urlcolor=NavyBlue]{hyperref}
\usepackage[all]{hypcap}
\title[Recent Star Formation in NGC~1052]{The most recent burst of Star Formation in the Massive Elliptical Galaxy NGC~1052\thanks{Based on European Southern Observatory (\textsf{ESO}) Very Large Telescope (VLT) program \mbox{076.B-0493} and \emph{Hubble Space Telescope} (\emph{HST}) programs ID 3639, 6286, 7403 and 7886.}}

\author[Fern\'andez-Ontiveros et~al.]{J.~A. Fern\'andez-Ontiveros$^{1,2}$\thanks{E-mail: \sf jafo@iac.es}, C. L\'opez-Sanjuan$^3$, M. Montes$^{1,2}$, M.~A. Prieto$^{1,2}$ and \newauthor J.~A. Acosta-Pulido$^{1,2}$\\
$^1$Instituto de Astrof\'isica de Canarias (IAC), V\'ia L\'actea s/n, La Laguna, E--38200, Spain\\
$^2$Departamento de Astrof\'isica, Facultad de F\'isica, Universidad de La Laguna, Astrof\'isico Fco. S\'anchez s/n, La Laguna, E--38207, Spain\\
$^3$Laboratoire d'Astrophysique de Marseille, P\^ole de l'Etoile Site de Ch\^ateau-Gombert 38, rue Fr\'ed\'eric Joliot-Curie, F--13388 Marseille, France}

\begin{document}

%\date{\today}
\date{Accepted 2010 November 9. Received 2010 October 8; in original form 2010 April 7}

\pagerange{\pageref{firstpage}--\pageref{lastpage}} \pubyear{2010}

\maketitle

\label{firstpage}

%%%%%%%%%%%%%%%%%%%%%%%%%%%%%%%%%%%%%%%%%%%%%%%%%%%%%%%%%%%%%%%%%%%%%%%
\begin{abstract}
High-spatial resolution near-infrared (NIR) images of the central $24$\,$\times$\,$24\, \rm{arcsec^2}$ ($\sim$\,$2$\,$\times$\,$2\, \rm{kpc^2}$) of the elliptical galaxy NGC~1052 reveal a total of 25 compact sources randomly distributed in the region. Fifteen of them exhibit H$\alpha$ luminosities an order of magnitude above the estimate for an evolved population of extreme horizontal branch stars. Their H$\alpha$ equivalent widths and optical-to-NIR spectral energy distributions are consistent with them being young stellar clusters aged $<$\,$7\, \rm{Myr}$. We consider this to be the first direct observation of spatially resolved star-forming regions in the central kiloparsecs of an elliptical galaxy. The sizes of these regions are $\lesssim$\,$11\, \rm{pc}$ and their median reddening is ${\rm E}(B - V)$\,$\sim$\,$1\, \rm{mag}$. According to previous works, NGC~1052 may have experienced a merger event about $1\, \rm{Gyr}$ ago. On the assumption that these clusters are spreaded with similar density over the whole galaxy, the fraction of galaxy mass ($5$\,$\times$\,$10^{-5}$) and rate of star formation ($0.01\, \rm{M_\odot\, yr^{-1}}$) involved, suggest the merger event as the possible cause for the star formation we see today.
\end{abstract}


%%%%%%%%%%%%%%%%%%%%%%%%%%%%%%%%%%%%%%%%%%%%%%%%%%%%%%%%%%%%%%%%%%%%%%%

\section{Introduction}

About $\sim$\,30\% of the local early-type galaxies present signs of Recent Star Formation (RSF) in the last Gyr \citep{2007ApJS..173..619K}. This result arises from the study of the integrated UV--optical colours of a large sample of early-type galaxies. A sizeable fraction of objects are revealed by their blue colours, \mbox{\emph{NUV}$ - r$\,$<$\,$5.5\, \rm{mag}$}, despite of the fact that their optical colours place them in the optical ``red sequence'' \citep[$u - r$\,$>$\,$2.2\, \rm{mag}$,][]{2001AJ....122.1861S}. Given these blue \emph{NUV}$- r$ colours, a merger is likely the origin of the induced RSF \citep{2009MNRAS.394.1713K,2010MNRAS.tmp.1161K,2010arXiv1009.5921L}.

In that scenario, one would expect the formation of new young cluster populations in an elliptical galaxy which presents blue UV--optical colours and a past merger event \citep{2010MNRAS.tmp.1161K}. However, most of the examples reported up to date exhibit intermediate--age clusters, between $\gtrsim$\,$200\, \rm{Myr}$ and a few $\rm{Gyr}$ \citep[e.g. NGC~1316, NGC~1380, NGC~1700, NGC~3610, NGC~5128][]{1999AJ....117.1700C,2001MNRAS.322..643G,2010ApJ...716...71W}. In those cases, most of the age estimates are based on the comparison of cluster colours with stellar population models, and thus are strongly dependent on the metallicity, extinction and IMF assumed in these models. Therefore, a scrutinized study of nearby ellipticals with blue UV--optical colours is particularly interesting to reveal the young cluster population at the earliest stage of their evolution, i.e. $\lesssim$\,$10\, \rm{Myr}$. It is in this context that this letter is addressed. It presents the first detection of individually resolved young stellar clusters, $\lesssim$\,$7\, \rm{Myr}$ old, in a nearby elliptical galaxy, NGC~1052. To single out this population, high spatial resolution imaging from UV to optical and infrared (IR) are used. A complete spectral energy distribution (SED) is built for each of the clusters, their nature, ages and dust estimates have been addressed on the basis of their H$\alpha$ emission.
\begin{figure*}
  \includegraphics[width=\textwidth]{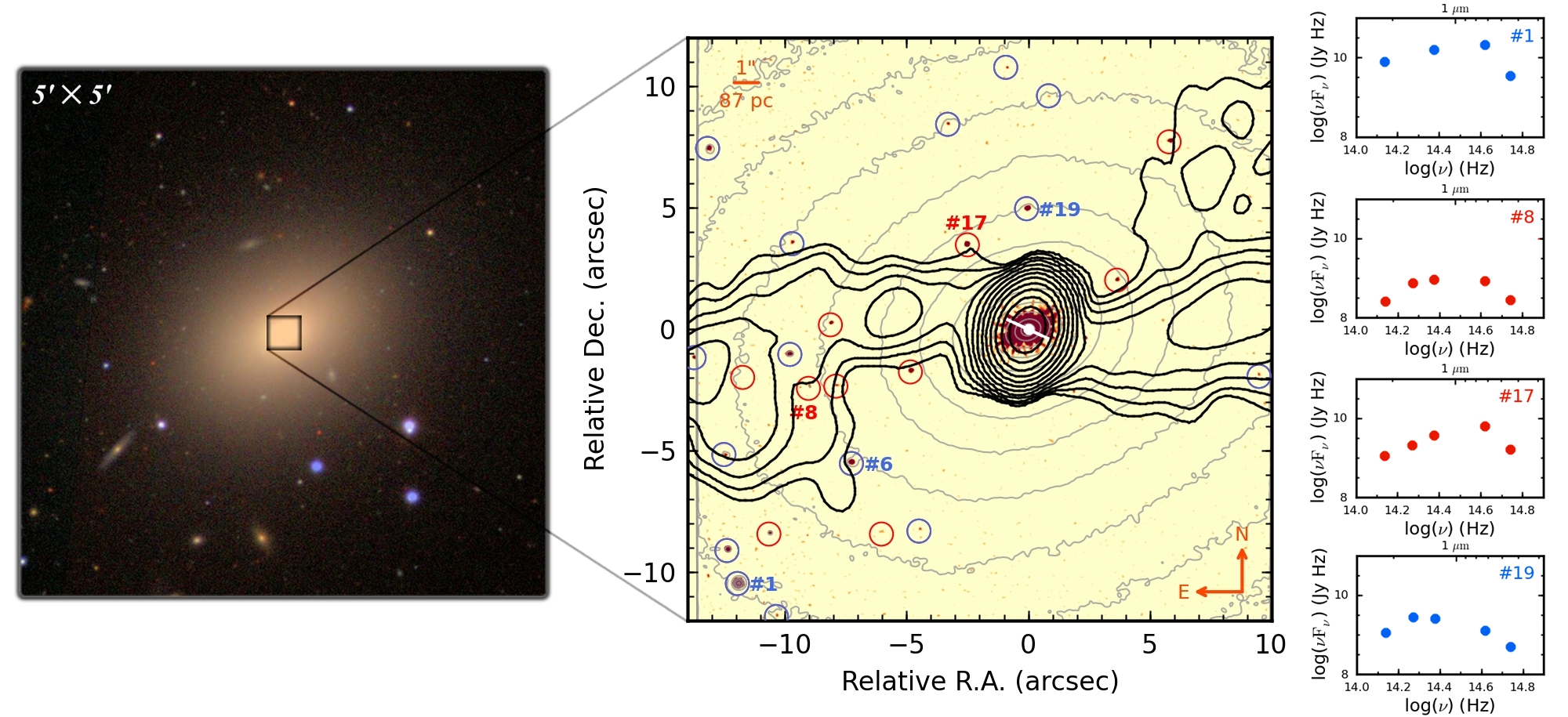}
  \protect\caption[NGC 1052 unsharp-masked image]{\emph{Left:} SDSS \emph{g} (blue), \emph{r} (green) and \emph{i}-band (red) colour composite image for NGC~1052. VLT/\textsf{NaCo} $24$\,$\times$\,$24\, \rm{arcsec^2}$ FOV is marked by a black square. \emph{Centre:} Unsharp-masked VLT/\textsf{NaCo} \emph{Ks}-band image, logarithmic contours for the natural \emph{Ks}-band image (grey colour) and VLA contours at $1.4\, \rm{GHz}$ \citep[black, beam $\sim$\,$2''$\,$\times$\,$1\farcs5$,][]{2007ApJS..171..376C}. In blue, knots with EWH$\alpha$\,$\ge$\,$50$\,\AA, in red those with $<$\,$50$\,\AA. The central white barred-dot marks the position of the nucleus and the direction of the parsec-scale twin jet. \emph{Right:} SEDs for two ``blue'' (\#1 and \#19) and two ``red'' knots (\#8 and \#17).}\label{K_med}
\end{figure*}

NGC~1052 is one of the nearest elliptical galaxies \citep[E4, $18\, \rm{Mpc}$, $1''$\,$\approx$\,$87\, \rm{pc}$,][]{2003ApJ...583..712J} with a stellar mass of $M_\star$\,$\approx$\,$10^{11}\, \rm{M_\odot}$ (derived from 2MASS\footnote{2 Micron All Sky Survey, see \url{http://www.ipac.caltech.edu/2mass/}} photometry, assuming $M_{\star}/L_{\rm{K}} = 1.32\, \rm{M_\odot/L_\odot}$, \citealt{2001MNRAS.326..255C}) and a low-level activity LINER nucleus \citep{1980A&A....87..152H}. It presents two radio lobes \citep[][Fig.~\ref{K_med}]{2007ApJS..171..376C} and a collimated ionized gas structure in H$\alpha$+[\textsc{N\,ii}] \citep{2000ApJ...532..323P}, both extended in the east--west direction, and a parsec-scale twin jet (PA\,$\sim$\,$60^\circ$) with an emission gap between the brightest part located north-east (NE) and the counter jet in the south-west (SW) direction \citep{2004A&A...420..467K}. The radio lobes are also spatially coincident with the X-ray emission region reported by \citet{2004A&A...420..467K}. The distribution of the \textsc{H\,i} gas, with two tidal tails extended from NE to SW and also an infalling component, is interpreted as evidence of a merger event about $1\, \rm{Gyr}$ ago \citep{1986AJ.....91..791V,1989AJ.....97..708V}. This scenario is further supported by the different rotation axes found for the gas and the stars \citep{1996A&A...307..391P} and by the presence of dust lanes along the N--S direction near the galaxy centre \citep{1990NASCP3098..431F}. Furthermore, NGC~1052 shows typical integrated colours of a ``rejuvenated'' early-type galaxy \citep[following][]{2007ApJS..173..619K}, having \mbox{\emph{NUV}$- r = 4.9\, \rm{mag}$}, while the \mbox{$u - r = 2.8\, \rm{mag}$} colour places this galaxy in the optical red sequence (\emph{NUV} from \citealt{2007MNRAS.381..245R}; \emph{u} and \emph{r}-band from SDSS\footnote{Sloan Digital Sky Survey, see \url{http://www.sdss.org/}}).

%%%%%%%%%%%%%%%%%%%%%%%%%%%%%%%%%%%%%%%%%%%%%%%%%%%%%%%%%%%%%%%%%%%%%%%

\section[]{Observations}\label{obs}

The brightness of the LINER nucleus permitted us to perform high-spatial resolution observations using the Very Large Telescope (VLT) and the Nasmyth Adaptive Optics System (NAOS) plus the Near-Infrared Imager and Spectrograph (CONICA). The VLT/\textsf{NaCo} dataset consist of near-infrared (NIR) \emph{J}, \emph{Ks} and \emph{L$'$}-band observations for the central $24$\,$\times$\,$24\, \rm{arcsec^2}$ of NGC~1052 ($\sim$\,$2$\,$\times$\,$2\, \rm{kpc^2}$, Fig.~\ref{K_med}), taken on 2005 December 3. The total exposure times are 900\,s (\emph{J}), 400\,s (\emph{Ks}) and 5.25\,s (\emph{L$'$}), and the achieved \textsc{fwhm} resolution --measured in the most compact objects in these images-- are $\lesssim$\,$0\farcs26$, $\lesssim$\,$0\farcs12$ and $\lesssim$\,$0\farcs12$, respectively. Data reduction of the \textsf{NaCo} dataset was performed using the \textsc{eclipse} package, provided by \textsf{ESO}, and includes sky subtraction, registration and combination of frames to obtain a final image per filter. The photometric calibration was based on standard stars taken along with the science frames. The relative photometric error is $\sim$\,$8\%$, $\sim$\,$11\%$ and $\sim$\,$11\%$ in the \emph{J}, \emph{Ks} and \emph{L$'$}-bands, respectively. The VLT/\textsf{NaCo} dataset was complemented with the following \emph{Hubble Space Telescope} (\emph{HST}) imaging: WFPC/F555W (Wide Field Planetary Camera/\emph{V}-band, $\lambda$\,$0.55\, \rm{\micron}$ $\Delta \lambda$\,$0.05\, \rm{\micron}$), WFPC2/F658N (including H$\alpha$+[\textsc{N\,ii}], $\lambda$\,$0.659\, \rm{\micron}$ $\Delta \lambda$\,$0.003\, \rm{\micron}$), STIS/F28X50LP (Space Telescope Imaging Spectrograph, $\lambda$\,$0.72\, \rm{\micron}$ $\Delta \lambda$\,$0.11\, \rm{\micron}$) and NICMOS/F160W (Near-Infrared Camera and Multi-Object Spectrometer/\emph{H}-band, $\lambda$\,$1.61\, \rm{\micron}$ $\Delta \lambda$\,$0.12\, \rm{\micron}$). The reduced images were retrieved directly from the scientific archive, and the photometric calibration was performed by using the \textsc{photflamb} keyword value in the image header as a conversion factor.

Image alignment of the whole set is based on the identification of two compact sources (knots) seen in all filters but in the \emph{L$'$}-band. We avoided using the nucleus since dust obscuration may distort its position in the optical. For the \emph{L$'$}-band, we assumed its position to match that in the \emph{Ks}-band. We further used a $1.4\, \rm{GHz}$ image with a beam resolution of $\sim$\,$2''$\,$\times$\,$1\farcs5$ from VLA \citep{2007ApJS..171..376C}. The signal-to-noise achieved in the \textsf{NaCo} \emph{J} and \emph{Ks}-bands permitted us to distinguish a number of point-like sources in the FOV. In order to enhance the image contrast, different techniques were applied, the best results being obtained by the unsharp-masking method \citep{1993PASP..105..308S}. A median circular filter with radius $0\farcs12$ was used to smooth the images, which were then subtracted from the original ones. The result for the \emph{Ks}-band is shown in the background of Fig.~\ref{K_med}. The \textsc{daophot} finding algorithm \citep{1987PASP...99..191S} recovered, on the \emph{J} and \emph{Ks} unsharp-masked images, a total of 25 point-like sources over a $3\sigma$ threshold in the central $24$\,$\times$\,$24\, \rm{arcsec^2}$ (circles in Fig.~\ref{K_med}). None of these knots are detected in the UV images from \emph{HST}/ACS (Advance Camera for Surveys), acquired with F250W and F330W filters and taken also from the scientific archive. The wider FOV of STIS image ($\sim$\,$28$\,$\times$\,$51\, \rm{arcsec^2}$) shows nearly 70 knots, which seem more dispersed with increasing radius. Counterparts in other bands (F555W, \emph{J}, \emph{Ks}, none detected in the \emph{L$'$}-band) exist only for the 25 knots in the common FOV. In the \emph{H}-band 2 of these knots are outside of the NICMOS FOV. The photometry was extracted from a circular aperture ($r_a$\,$=$\,\textsc{fwhm} in each filter) centred on a common position for all images. The background emission was then subtracted as the mode value for an annular concentric region (\mbox{$r_s$\,$\sim$\,$[0\farcs4$--$0\farcs6]$}). Some of these regions stand out in the narrow-band F658N image, centred on H$\alpha$+[\textsc{N\,ii}]. To estimate the flux in this blend, the continuum contribution was inferred by linear interpolation between the F555W and \emph{J} broad-band filters. The possible uncertainties derived in the continuum estimation will be discussed later. Finally, the H$\alpha$ flux was then corrected by the [\textsc{N\,ii}] contribution by assuming that to be 40 per cent of the total \citep{1981PASP...93....5B}.

%%%%%%%%%%%%%%%%%%%%%%%%%%%%%%%%%%%%%%%%%%%%%%%%%%%%%%%%%%%%%%%%%%%%%%%

\section[]{Results}\label{res}

Most of the 25 knots are unresolved or barely resolved with sizes of about $0\farcs12$ ($\lesssim$\,$10.5\, \rm{pc}$), i.e. the spatial resolution in the VLT/\textsf{NaCo} \emph{Ks}-band image. They show a median brightness of $m_K$\,$=$\,$20.9\, \rm{mag}$, which corresponds to an absolute magnitude of $M_K$\,$=$\,$-10.4\, \rm{mag}$, too bright for being individual red supergiant stars \citep{2009ApJ...703L..72T}. The knots are distributed in the $-12$\,$\le$\,$M_K$\,$\le$\,$-9\, \rm{mag}$ range, except knot~\#1, which is by far the brightest with $M_{K_{\#1}}$\,$=$\,$-13.6\, \rm{mag}$. In the optical, they are characterized by a median $m_V$\,$=$\,$24.3\, \rm{mag}$ and $m_I$\,$=$\,$22.1\, \rm{mag}$, corresponding to $M_V$\,$=$\,$-7.0\, \rm{mag}$ and $M_I$\,$=$\,$-9.2\, \rm{mag}$. Most of the 25 knots are strong emitters in H$\alpha$. Moreover, the H$\alpha$ equivalent width (EWH$\alpha$) distribution shows $\gtrsim$\,$50$\,\AA \ for most of them, we therefore adopted this value as a threshold to separate H$\alpha$ emitting knots from those without H$\alpha$ emission. Assuming these to be star clusters, their ages were derived from their EWH$\alpha$. \textsc{starburst99} models \citep{1999ApJS..123....3L} were used, assuming an instantaneous burst model with the highest available metallicity, $Z$\,$=$\,$0.04$. This follows the results by \citet{2005MNRAS.358..419P} who found higher than solar metallicity abundances ($Z$\,$>$\,$0.08$) in the central $\sim$\,$10\, \rm{arcsec}$ ($\sim$\,$1\, \rm{kpc}$) region of NGC~1052 from the analysis of spectral indices measured at various radii in an optical long-slit spectrum. Notice that these are stellar metallicities and not nebular ones. A lower limit for the metallicity will be considered below.

\emph{A total of 15 knots were found with EWH$\alpha$\,$\ge$\,$50$\,\AA \ (blue circles in Fig.~\ref{K_med}), which yields ages of $\lesssim$\,$6.8\, \rm{Myr}$}. Their median EWH$\alpha$ value is $116$\,\AA \ ($6.2\, \rm{Myr}$), with half of them located in the $105$--$175$\,\AA \ range ($5.4$--$6.5\, \rm{Myr}$), i.e.\ the first and third quartiles, respectively. Lower EWH$\alpha$ for the remaining knots may be due to the clusters being older or to a higher extinction which will affect the H$\alpha$ continuum interpolation. In the conservative side, choosing larger photometric aperture in the F555W and \emph{J}-bands --the ones used to define the H$\alpha$ continuum flux-- decreases the number of knots over the $50$\,\AA \ limit to 8, instead of 15. The net effect of considering lower metallicities, e.g. $Z$\,$=$\,$0.001$, would translate in older ages for a given EWH$\alpha$, but values over the $50$\,\AA \ threshold still limit the age to $\le$\,$12.2\, \rm{Myr}$. The number of ``young'' regions just increases to 16 by assuming the [\textsc{N\,ii}] contribution to be 20 per cent instead of 40 in the H$\alpha$+[\textsc{N\,ii}] blend.

The H$\alpha$ luminosity ($L_{\rm{H\alpha}}$) for the 15 knots with EWH$\alpha$\,$\ge$\,$50$\,\AA \ spans the $[1$--$33]$\,$\times$\,$10^{36}\, \rm{erg\, s^{-1}}$ range. Notwithstanding, ionizing photons may be produced by an old stellar population due to the ``UV--upturn'' \citep[see][and references therein]{1999ARA&A..37..603O}. This additional contribution in the UV range originates in some low-mass, helium-burning stars in the extreme horizontal branch (EHB) and subsequent phases of evolution. In this scenario, the compact knots in NGC~1052 could be either old globular clusters (GCs) with an EHB component or young stellar clusters probably linked with the merger event occurred in the recent history of this galaxy. A detailed analysis was then performed for knot~\#6 (Fig.~\ref{K_med}), which is representative of the 15 knots' sample in terms of brightness and SED's shape. From its H$\alpha$ luminosity ($5.2$\,$\times$\,$10^{36}\, \rm{erg\, s^{-1}}$) we derived an ionizing photon rate of $\sim$\,$3.8$\,$\times$\,$10^{48}\, \rm{ph\, s^{-1}}$ \citep[following]{1989agna.book.....O}. This estimate is compared in the next sections with predictions of a SED from an old and a young stellar population, respectively.

\subsection{Old globular clusters?}\label{GC}
\begin{figure}
  \includegraphics[width=\columnwidth]{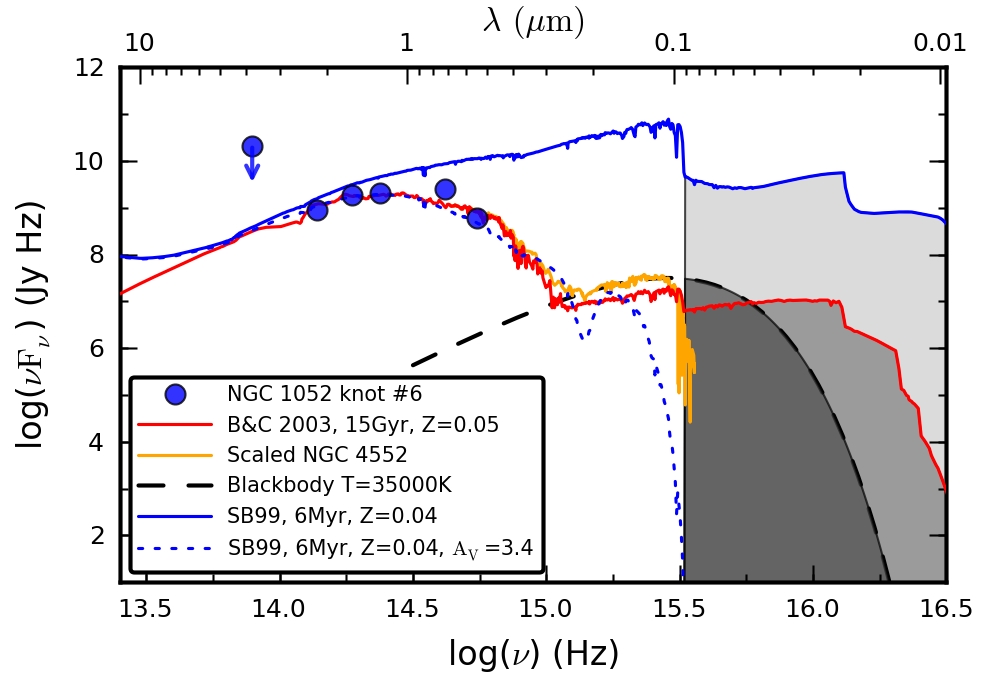}
  \protect\caption[Ultraviolet upturn estimation]{Knot~\#6 SED (blue circles) compared with a $15\, \rm{Gyr}$, $Z$\,$=$\,$0.05$ template \citep[][red line]{2003MNRAS.344.1000B}, a composite spectrum of NGC~4552 \citep[][yellow line]{1998ApJ...492..480Y} plus a blackbody spectrum ($T$\,$=$\,$35000\, \rm{K}$, black-dashed line) and a $6\, \rm{Myr}$, $Z$\,$=$\,$0.04$ template from \textsc{Starburst99} \citep[][blue-solid line]{1999ApJS..123....3L}, also shown with an obscuration of $A_V$\,$=$\,$3.4\, \rm{mag}$ (blue-dotted line). Shaded areas correspond to photons with $\lambda$\,$\le$\,$910$\,\AA.}\label{n1052_uvx}
\end{figure}

Fig.~\ref{n1052_uvx} shows the spectral energy distribution (SED) of the knot~\#6, which is derived from broad-band photometry (blue dots) including the upper limit in the \emph{L}-band. First, we show the SED of a $15\, \rm{Gyr}$ old, $Z$\,$=$\,$0.05$ population from \citet[][red line in Fig.~\ref{n1052_uvx}]{2003MNRAS.344.1000B} scaled to the NIR SED of knot~\#6 (the NIR SED comprises 3 points: \emph{HST}/F160W, \textsf{NaCo} \emph{J} and \emph{Ks}-bands). We choose scaling to the NIR to minimize dust extinction effects. The rate of ionizing photons produced by this population --mainly by blue horizontal branch stars-- is then determined by integrating the template below the Lyman limit (shaded area below the red line). This results in $1.2$\,$\times$\,$10^{47}\, \rm{ph\, s^{-1}}$, i.e. more than an order of magnitude lower than the value measured for knot~\#6; thus, the H$\alpha$ emission cannot be explained by this $15\, \rm{Gyr}$ old population.

However, the EHB component is not included in the $15\, \rm{Gyr}$ old template. EHB stars present small envelopes around the helium core, showing a hot thermal spectrum with $T$\,$\gtrsim$\,$20000\, \rm{K}$. The integrated energy distribution of such an old population ($\gtrsim$\,$10\, \rm{Gyr}$) increases shortwards of $\sim$\,$2000$\,\AA \ and is thus capable of emitting FUV photons \citep{1990ApJ...364...35G}. To estimate the possible contribution of these stars we consider the spectrum of the UV-strong elliptical galaxy NGC~4552, which is expected to have an important contribution from EHB stars. For consistency, this spectrum was scaled to the $15\, \rm{Gyr}$ old template used before in the optical region (yellow line in Fig~\ref{n1052_uvx}). Shortwards of $900$\,\AA, NGC~4552 spectrum was extended with a $T$\,$=$\,$35000\, \rm{K}$ black body that was scaled to NGC~4552 in the UV part (black-dashed line in Fig.~\ref{n1052_uvx}). In this case, the integrated number of ionizing photons is $1.8$\,$\times$\,$10^{47}\, \rm{ph\, s^{-1}}$, still an order of magnitude below the measured value. It should be noticed that the the measured H$\alpha$ emission in the knots, and thus the ionizing photon rate, may still be an lower limit due to the possible contribution of H$\alpha$ in absorption. This is expected from intermediate-age A-stars population in the clusters.

\subsection{Young stellar clusters?}\label{YSC}
We now consider a young $6\, \rm{Myr}$ old, $Z$\,$=$\,$0.04$ star cluster model from \textsc{Starburst99} \citep[blue-solid line in Fig.~\ref{n1052_uvx},][]{1999ApJS..123....3L}. The model choice is based on the EWH$\alpha$ of knot~\#6. Scaling it to the NIR SED region of this knot, the resulting photon rate is $2.5$\,$\times$\,$10^{49}\, \rm{ph\, s^{-1}}$. This is more than an order of magnitude larger than the measured value. Moreover, the model largely overestimates the emission in the optical range (see Fig.~\ref{n1052_uvx}). As dust is expected in the environment of young clusters, different extinction values were applied to this model till matching the observed both SED and H$\alpha$ luminosity. We started with an initial value of $A_V$\,$=$\,$2.5\, \rm{mag}$ --inferred from the H$\alpha$ ratio between model and observation-- and apply it consecutively to the SED. The process of fitting both H$\alpha$ emission and SED shape was done in an iterative manner \citep[using the extinction curve by][]{1989ApJ...345..245C}. The best match was found for a model with $A_V = 3.4\, \rm{mag}$ (blue-dotted line in Fig.~\ref{n1052_uvx}), which reproduces the shape of the measured SED. The intrinsic photon rate inferred from the model is $4.9$\,$\times$\,$10^{49}\, \rm{ph\, s^{-1}}$, $L_{\rm{H\alpha}}$\,$\sim$\,$6.6\times$\,$10^{37}\, \rm{erg\, s^{-1}}$ and mass $1.3$\,$\times$\,$10^4\, \rm{M_\odot}$ \citep[from \emph{Ks}-band luminosity, assuming an age--$M/L_K$ dependence from][]{1999ApJS..123....3L}. Note that, as we are assuming a foreground dust screen, these values may be lower limits.

\subsection{Colours}
\begin{figure}
  \includegraphics[width=\columnwidth]{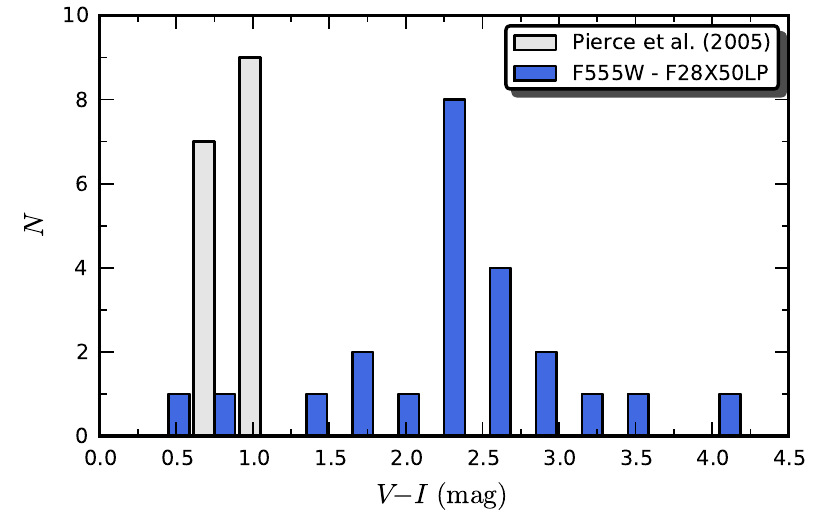}
  \protect\caption[V - I colour]{In grey, $V - I$ colours for the GCs in the outer $2$--$20\, \rm{kpc}$ of NGC~1052 \citep{2005MNRAS.358..419P}. In blue, the F555W$-$F28X50LP colour for the 25 knots identified in the central $\sim$\,$2$\,$\times$\,$2\, \rm{kpc^2}$.}\label{n1052_VI}
\end{figure}

All the 25 knots look different in brightness and colours with respect to those of the GCs studied in the outer $2$--$20\, \rm{kpc}$ of this galaxy by \citet{2005MNRAS.358..419P}. The latter present a median $m_V$\,$\sim$\,$22.0\, \rm{mag}$ with $V - I$\,$\sim$\,$0.9\, \rm{mag}$, in contrast with the $V - I$\,$=$\,$2.4\, \rm{mag}$\footnote{Strictly speaking, F555W$-$F28X50LP colour.} shown by the former (Fig.~\ref{n1052_VI}). Note the \emph{HST}/F28X50LP filter is even bluer than the \emph{I}-band. The differences suggest that both populations are intrinsically distinct. This filter also includes the H$\alpha$+[\textsc{N\,ii}] blend ($\sim$\,$0.660\, \rm{\micron}$ at $z$\,$=$\,$0.005$) very close to the transmission peak\footnote{\url{http://www.stsci.edu/hst/stis/design/filters/}}, but this contribution to the broad-band magnitude ($\sim$\,$0.3\, \rm{dex}$ for knot~\#6) is not enough to explain the large difference in $V - I$ colour. Only knot~\#1 has a $m_V$\,$=$\,$21.9\, \rm{mag}$, compatible with that of the GCs, but its $V - I$ colour is much redder ($2.6\, \rm{mag}$).

This discrepancy between the two populations cannot be ascribed to the colour bimodality reported by \citet{2001MNRAS.325.1431F}, since all the GCs in this study lie in the range $1.2$\,$<$\,$V - I$\,$<$\,$1.7\, \rm{mag}$. On the other hand, an obscuration of $A_V$\,$=$\,$3.4\, \rm{mag}$ may explain why the knots in the central region look much redder than those of the GCs. Similar extinction values and colours to those shown by the knots have been reported for young stellar clusters (YSCs) in the starbursts galaxies NGC~5253 \citep{2004A&A...415..509V} and NGC~253 \citep{2009MNRAS.392L..16F}.

%%%%%%%%%%%%%%%%%%%%%%%%%%%%%%%%%%%%%%%%%%%%%%%%%%%%%%%%%%%%%%%%%%%%%%%

\section[]{Discussion}\label{discuss}

Recent star formation is found in the centre ($2$\,$\times$\,$2\, \rm{kpc^2}$) of NGC~1052 in the form of young ($\sim$\,$6\, \rm{Myr}$) and compact ($\sim$\,$14\, \rm{pc}$) stellar clusters with masses of $\sim$\,$10^4\, \rm{M_\odot}$. The age estimation do not depends strongly on the stellar population model assumed. The total star formation rate (SFR) in the centre, derived from the H$\alpha$ luminosity of the 25 knots, is SFR\,$\sim$\,$7$\,$\times$\,$10^{-4}\, \rm{M_\odot\, yr^{-1}}$ \citep[following][]{1998ARA&A..36..189K}. Assuming that these clusters are spreaded with similar density over the whole galaxy results in an upper limit for the SFR of $\sim$\,$0.01\, \rm{M_\odot\, yr^{-1}}$. The resulting specific SFR places the galaxy in the ``red and death'' sequence \citep[SFR/$M_{\star}$\,$\lesssim$\,$10^{-12}\, \rm{yr^{-1}}$,][]{2009A&A...501...15F}. Even if considering the extrapolation for the whole galaxy, the total mass forming these clusters would be less than $5$\,$\times$\,$10^{-3}$ per cent of the galaxy stellar mass. This suggest that the current RSF is not dominant in the present evolution of NGC~1052.

Several observational signatures indicate that NGC~1052 experienced a merger in the last $1\, \rm{Gyr}$ (c.f. Introduction). The origin of the knots may be associated with left over activity from that event. Hydrodynamical \emph{N}-body merger simulations (mass ratio 1:6 to 1:10) by \citet{2010MNRAS.405.2327P} suggest that star formation occurs in several bursts after the merger event rather than being a continuous process \citep[see also][]{2009MNRAS.394.1713K}. This is in agreement with the observations: we find two very young clusters (EWH$\alpha$\,$\sim$\,$230$\,\AA; $\sim$\,$5.1\, \rm{Myr}$), 4 young ones ($140$\,$\le$\,EWH$\alpha$\,$<$\,$200$\,\AA; $5.2$--$5.6\, \rm{Myr}$), 7 slightly older ones ($90$\,$\le$\,EWH$\alpha$\,$<$\,$140$\,\AA; $5.6$--$6.6\, \rm{Myr}$) and 12 with EWH$\alpha$\,$<$\,$90$\,\AA \ ($>$\,$6.6\, \rm{Myr}$). This means that \emph{we are probably looking at either the final stage or a minimum in the star formation activity in the centre induced by a merger event}. In their simulations, \citet{2009MNRAS.394.1713K} explore extinction values in the ${\rm E}(B - V)$\,$=$\,$0 - 0.5\, \rm{mag}$ range, showing that observations favour those models with higher extinction (${\rm E}(B - V)$\,$\gtrsim$\,$0.3\, \rm{mag}$). For comparison, the colour excess of knot~\#6 suggest higher values, i.e.\ ${\rm E}(B - V)$\,$\sim$\,$1.1\, \rm{mag}$ ($A_V$\,$\sim$\,$3.4\, \rm{mag}$).

Regarding at their spatial distribution, the knots with EWH$\alpha$\,$\ge$\,$50$\,\AA \ are located randomly and do not seem to follow any special pattern except for a weak preference for the NW--SE direction. Moreover, we do not find any clear connection with either the kpc-scale radio lobes at $1.4\, \rm{GHz}$ (Fig.~\ref{K_med}) or with the extended emission in X-rays detected by \emph{Chandra} \citep{2004A&A...420..467K}. This indicates that the recent star-formation episode in the centre is not, at least directly, driven by nuclear activity.

The lack of high spatial resolution CO maps does not allow us to connect the knots with density enhancements in the cold material, but the dust lanes in the N--S direction seen in the STIS image are not apparently related with the knots. However, this scattered star formation is in line with the ``widespread'' star-formation mode found by \citet{2010MNRAS.402.2140S} for early-type galaxies with multiple and mismatched kinematic components. This is the case of NGC~1052, which shows 2 counter-rotating gaseous components decoupled from the stellar kinematics \citep{1996A&A...307..391P}. Furthermore, \citet{2010MNRAS.402.2140S} find that those objects with very low SFRs ($\lesssim$\,$0.06\, \rm{M_\odot\, yr^{-1}}$) exhibit a more centralized star-formation activity, which is also favoured by recent simulations \citep{2010MNRAS.405.2327P}. This might explain why \citet{2001MNRAS.325.1431F} and \citet{2005MNRAS.358..419P} did not find young clusters in their study, as they covered the outskirts of NGC~1052 ($2$--$20\, \rm{kpc}$). Nevertheless, an H$\alpha$ image with a wider FOV is needed to check the young cluster distribution on scales larger than the inner few kpc.

Since we determine the age of the knots on the basis of H$\alpha$ emission, we are only sensitive to very recent star-formation episodes ($\lesssim$\,$10\, \rm{Myr}$) and cannot provide a reliable age for those knots with EWH$\alpha$\,$<$\,$50$\,\AA \ (10 out of the 25). However, these show very similar properties (i.e.\ sizes, magnitudes, colours) to those with EWH$\alpha$\,$>$\,$50$\,\AA, are also randomly distributed and their SEDs are qualitatively similar (see Fig.~\ref{K_med}). Results from stellar population models on the integrated spectrum of the central $\sim$\,$1\, \rm{kpc}$ indicate luminosity-weighted ages of $\sim$\,$2\, \rm{Gyr}$ \citep{2005MNRAS.358..419P}. However, the contribution of the 25 knots reported in this work is not sufficient to ``rejuvenate'' the integrated emission of the central region. In line with the scenario proposed by \citet{2010MNRAS.402.2140S}, a middle--age population ($10\, \rm{Myr}$\,$<$\,$t$\,$<$\,$2\, \rm{Gyr}$) is still needed, that would be probably related with a more intense star-formation episode in the past. A spatially resolved spectroscopic study of the central kpc region would be enlightening.

%%%%%%%%%%%%%%%%%%%%%%%%%%%%%%%%%%%%%%%%%%%%%%%%%%%%%%%%%%%%%%%%%%%%%%%

\section{Final Remarks}\label{remarks}

We propose that the knots found in the centre of NGC~1052 are young stellar clusters (YSCs) formed in a very recent star formation episode, probably related with the merger event that occurred $\sim$\,$1\, \rm{Gyr}$ ago \citep{1986AJ.....91..791V}. This is the first detection of resolved YSCs in the centre of an elliptical galaxy. Their presence has three major implications: \emph{i)} we are looking at the final stage of the RSF occurring in the central parsecs of NGC~1052, \emph{ii)} the RSF takes place in the form of compact and young stellar clusters similar to those found in starburst galaxies, and \emph{iii)} the reddening predicted by young stellar cluster templates [\mbox{${\rm E}(B - V)$\,$\sim$\,$1.1\, \rm{mag}$}] should be explored in future simulations of these systems, which used to assume a lower value as an upper limit.

The case of NGC~1052 is probably one of the best examples for the study of the recent evolution of nearby early-type galaxies via merging with gas-rich satellites. However, high spatial resolution observations in dynamically younger remnants, for which we expect a larger number of YSCs, are desired to understand in detail the role of mergers in the present evolution of elliptical galaxies.

%%%%%%%%%%%%%%%%%%%%%%%%%%%%%%%%%%%%%%%%%%%%%%%%%%%%%%%%%%%%%%%%%%%%%%%
\footnotesize
\section*{Acknowledgments}
\textsl{The authors acknowledge A. Mar\'in-Franch for his useful hints on globular cluster physics and M. Orienti for reviewing the radio data. We also thank the
referee for detailed comments which helped to improve the original manuscript. This work is partially funded by the Spanish MEC project AYA2007--60235.}

%%%%%%%%%%%%%%%%%%%%%%%%%%%%%%%%%%%%%%%%%%%%%%%%%%%%%%%%%%%%%%%%%%%%%%%
\bibliography{ngc1052}

\begin{thebibliography}{}

\bibitem[\protect\citeauthoryear{{Baldwin}, {Phillips} \&
  {Terlevich}}{{Baldwin} et~al.}{1981}]{1981PASP...93....5B}
{Baldwin} J.~A.,  {Phillips} M.~M.,    {Terlevich} R.,  1981, \pasp, 93, 5

\bibitem[\protect\citeauthoryear{{Bruzual} \& {Charlot}}{{Bruzual} \&
  {Charlot}}{2003}]{2003MNRAS.344.1000B}
{Bruzual} G.,  {Charlot} S.,  2003, \mnras, 344, 1000

\bibitem[\protect\citeauthoryear{{Cardelli}, {Clayton} \& {Mathis}}{{Cardelli}
  et~al.}{1989}]{1989ApJ...345..245C}
{Cardelli} J.~A.,  {Clayton} G.~C.,    {Mathis} J.~S.,  1989, \apj, 345, 245

\bibitem[\protect\citeauthoryear{{Carlson}, {Holtzman}, {Grillmair}, {Mould},
  {Griffiths}, {Ballester}, {Burrows}, {Clarke}, {Crisp}, {Evans}, {Gallagher}
  III, {Hester}, {Hoessel}, {Scowen} \& {Stapelfeldt}}{{Carlson}
  et~al.}{1999}]{1999AJ....117.1700C}
{Carlson} M.~N.,  {Holtzman} J.~A.,  {Grillmair} C.~J.,  {Mould} J.~R.,
  {Griffiths} R.~E.,  {Ballester} G.~E.,  {Burrows} C.~J.,  {Clarke} J.~T.,
  {Crisp} D.,  {Evans} R.~W.,  {Gallagher} III J.~S.,  {Hester} J.~J.,
  {Hoessel} J.~G.,  {Scowen} P.~A.,    {Stapelfeldt} K.~R.,  1999, \aj, 117,
  1700

\bibitem[\protect\citeauthoryear{{Cole}, {Norberg}, {Baugh}, {Frenk},
  {Bland-Hawthorn}, {Bridges}, {Cannon}, {Colless}, {Collins}, {Couch} \& et
  al.}{{Cole} et~al.}{2001}]{2001MNRAS.326..255C}
{Cole} S.,  {Norberg} P.,  {Baugh} C.~M.,  {Frenk} C.~S.,  {Bland-Hawthorn} J.,
   {Bridges} T.,  {Cannon} R.,  {Colless} M.,  {Collins} C.,  {Couch} W.,    et
  al. 2001, \mnras, 326, 255

\bibitem[\protect\citeauthoryear{{Cooper}, {Lister} \& {Kochanczyk}}{{Cooper}
  et~al.}{2007}]{2007ApJS..171..376C}
{Cooper} N.~J.,  {Lister} M.~L.,    {Kochanczyk} M.~D.,  2007, \apjs, 171, 376

\bibitem[\protect\citeauthoryear{{Fern{\'a}ndez-Ontiveros}, {Prieto} \&
  {Acosta-Pulido}}{{Fern{\'a}ndez-Ontiveros}
  et~al.}{2009}]{2009MNRAS.392L..16F}
{Fern{\'a}ndez-Ontiveros} J.~A.,  {Prieto} M.~A.,    {Acosta-Pulido} J.~A.,
  2009, \mnras, 392, L16

\bibitem[\protect\citeauthoryear{{Fontana}, {Santini}, {Grazian}, {Pentericci},
  {Fiore}, {Castellano}, {Giallongo}, {Menci}, {Salimbeni}, {Cristiani},
  {Nonino} \& {Vanzella}}{{Fontana} et~al.}{2009}]{2009A&A...501...15F}
{Fontana} A.,  {Santini} P.,  {Grazian} A.,  {Pentericci} L.,  {Fiore} F.,
  {Castellano} M.,  {Giallongo} E.,  {Menci} N.,  {Salimbeni} S.,  {Cristiani}
  S.,  {Nonino} M.,    {Vanzella} E.,  2009, \aap, 501, 15

\bibitem[\protect\citeauthoryear{{Forbes}, {Georgakakis} \& {Brodie}}{{Forbes}
  et~al.}{2001}]{2001MNRAS.325.1431F}
{Forbes} D.~A.,  {Georgakakis} A.~E.,    {Brodie} J.~P.,  2001, \mnras, 325,
  1431

\bibitem[\protect\citeauthoryear{{Forbes}, {Sparks} \& {Macchetto}}{{Forbes}
  et~al.}{1990}]{1990NASCP3098..431F}
{Forbes} D.~A.,  {Sparks} W.~B.,    {Macchetto} F.~D.,  1990, NASA Conference
  Publication, 3098, 431

\bibitem[\protect\citeauthoryear{{Goudfrooij}, {Mack}, {Kissler-Patig},
  {Meylan} \& {Minniti}}{{Goudfrooij} et~al.}{2001}]{2001MNRAS.322..643G}
{Goudfrooij} P.,  {Mack} J.,  {Kissler-Patig} M.,  {Meylan} G.,    {Minniti}
  D.,  2001, \mnras, 322, 643

\bibitem[\protect\citeauthoryear{{Greggio} \& {Renzini}}{{Greggio} \&
  {Renzini}}{1990}]{1990ApJ...364...35G}
{Greggio} L.,  {Renzini} A.,  1990, \apj, 364, 35

\bibitem[\protect\citeauthoryear{{Heckman}}{{Heckman}}{1980}]{1980A&A....87..1%
52H}
{Heckman} T.~M.,  1980, \aap, 87, 152

\bibitem[\protect\citeauthoryear{{Jensen}, {Tonry}, {Barris}, {Thompson},
  {Liu}, {Rieke}, {Ajhar} \& {Blakeslee}}{{Jensen}
  et~al.}{2003}]{2003ApJ...583..712J}
{Jensen} J.~B.,  {Tonry} J.~L.,  {Barris} B.~J.,  {Thompson} R.~I.,  {Liu}
  M.~C.,  {Rieke} M.~J.,  {Ajhar} E.~A.,    {Blakeslee} J.~P.,  2003, \apj,
  583, 712

\bibitem[\protect\citeauthoryear{{Kadler}, {Kerp}, {Ros}, {Falcke}, {Pogge} \&
  {Zensus}}{{Kadler} et~al.}{2004}]{2004A&A...420..467K}
{Kadler} M.,  {Kerp} J.,  {Ros} E.,  {Falcke} H.,  {Pogge} R.~W.,    {Zensus}
  J.~A.,  2004, \aap, 420, 467

\bibitem[\protect\citeauthoryear{{Kaviraj}}{{Kaviraj}}{2010}]{2010MNRAS.tmp.11%
61K}
{Kaviraj} S.,  2010, \mnras, pp 1161--+

\bibitem[\protect\citeauthoryear{{Kaviraj}, {Peirani}, {Khochfar}, {Silk} \&
  {Kay}}{{Kaviraj} et~al.}{2009}]{2009MNRAS.394.1713K}
{Kaviraj} S.,  {Peirani} S.,  {Khochfar} S.,  {Silk} J.,    {Kay} S.,  2009,
  \mnras, 394, 1713

\bibitem[\protect\citeauthoryear{{Kaviraj}, {Schawinski}, {Devriendt},
  {Ferreras}, {Khochfar}, {Yoon}, {Yi}, {Deharveng}, {Boselli}, {Barlow} \& et
  al.}{{Kaviraj} et~al.}{2007}]{2007ApJS..173..619K}
{Kaviraj} S.,  {Schawinski} K.,  {Devriendt} J.~E.~G.,  {Ferreras} I.,
  {Khochfar} S.,  {Yoon} S.,  {Yi} S.~K.,  {Deharveng} J.,  {Boselli} A.,
  {Barlow} T.,    et al. 2007, \apjs, 173, 619

\bibitem[\protect\citeauthoryear{{Kennicutt}
  Jr.}{{Kennicutt}}{1998}]{1998ARA&A..36..189K}
{Kennicutt} Jr. R.~C.,  1998, \araa, 36, 189

\bibitem[\protect\citeauthoryear{{Leitherer}, {Schaerer}, {Goldader},
  {Gonz{\'a}lez Delgado}, {Robert}, {Kune}, {de Mello}, {Devost} \&
  {Heckman}}{{Leitherer} et~al.}{1999}]{1999ApJS..123....3L}
{Leitherer} C.,  {Schaerer} D.,  {Goldader} J.~D.,  {Gonz{\'a}lez Delgado}
  R.~M.,  {Robert} C.,  {Kune} D.~F.,  {de Mello} D.~F.,  {Devost} D.,
  {Heckman} T.~M.,  1999, \apjs, 123, 3

\bibitem[\protect\citeauthoryear{{L{\'o}pez-Sanjuan}
  et~al.,}{{L{\'o}pez-Sanjuan}  et~al.}{2010}]{2010arXiv1009.5921L}
{L{\'o}pez-Sanjuan} C.,  et~al., 2010, arXiv:1009.5921

\bibitem[\protect\citeauthoryear{{O'Connell}}{{O'Connell}}{1999}]{1999ARA&A..3%
7..603O}
{O'Connell} R.~W.,  1999, \araa, 37, 603

\bibitem[\protect\citeauthoryear{{Osterbrock}}{{Osterbrock}}{1989}]{1989agna.b%
ook.....O}
{Osterbrock} D.~E.,  1989, Astrophysics of gaseous nebulae and active galactic
  nuclei.
University Science Books

\bibitem[\protect\citeauthoryear{{Peirani}, {Crockett}, {Geen}, {Khochfar},
  {Kaviraj} \& {Silk}}{{Peirani} et~al.}{2010}]{2010MNRAS.405.2327P}
{Peirani} S.,  {Crockett} R.~M.,  {Geen} S.,  {Khochfar} S.,  {Kaviraj} S.,
  {Silk} J.,  2010, \mnras, 405, 2327

\bibitem[\protect\citeauthoryear{{Pierce}, {Brodie}, {Forbes}, {Beasley},
  {Proctor} \& {Strader}}{{Pierce} et~al.}{2005}]{2005MNRAS.358..419P}
{Pierce} M.,  {Brodie} J.~P.,  {Forbes} D.~A.,  {Beasley} M.~A.,  {Proctor} R.,
     {Strader} J.,  2005, \mnras, 358, 419

\bibitem[\protect\citeauthoryear{{Plana} \& {Boulesteix}}{{Plana} \&
  {Boulesteix}}{1996}]{1996A&A...307..391P}
{Plana} H.,  {Boulesteix} J.,  1996, \aap, 307, 391

\bibitem[\protect\citeauthoryear{{Pogge}, {Maoz}, {Ho} \& {Eracleous}}{{Pogge}
  et~al.}{2000}]{2000ApJ...532..323P}
{Pogge} R.~W.,  {Maoz} D.,  {Ho} L.~C.,    {Eracleous} M.,  2000, \apj, 532,
  323

\bibitem[\protect\citeauthoryear{{Rampazzo}, {Marino}, {Tantalo}, {Bettoni},
  {Buson}, {Chiosi}, {Galletta}, {Gr{\"u}tzbauch} \& {Rich}}{{Rampazzo}
  et~al.}{2007}]{2007MNRAS.381..245R}
{Rampazzo} R.,  {Marino} A.,  {Tantalo} R.,  {Bettoni} D.,  {Buson} L.~M.,
  {Chiosi} C.,  {Galletta} G.,  {Gr{\"u}tzbauch} R.,    {Rich} R.~M.,  2007,
  \mnras, 381, 245

\bibitem[\protect\citeauthoryear{{Shapiro} et~al.,}{{Shapiro}
  et~al.}{2010}]{2010MNRAS.402.2140S}
{Shapiro} K.~L.,  et~al., 2010, \mnras, 402, 2140

\bibitem[\protect\citeauthoryear{{Sofue}}{{Sofue}}{1993}]{1993PASP..105..308S}
{Sofue} Y.,  1993, \pasp, 105, 308

\bibitem[\protect\citeauthoryear{{Stetson}}{{Stetson}}{1987}]{1987PASP...99..1%
91S}
{Stetson} P.~B.,  1987, \pasp, 99, 191

\bibitem[\protect\citeauthoryear{{Strateva}, {Ivezi{\'c}}, {Knapp},
  {Narayanan}, {Strauss}, {Gunn}, {Lupton}, {Schlegel}, {Bahcall}, {Brinkmann}
  \& et al.}{{Strateva} et~al.}{2001}]{2001AJ....122.1861S}
{Strateva} I.,  {Ivezi{\'c}} {\v Z}.,  {Knapp} G.~R.,  {Narayanan} V.~K.,
  {Strauss} M.~A.,  {Gunn} J.~E.,  {Lupton} R.~H.,  {Schlegel} D.,  {Bahcall}
  N.~A.,  {Brinkmann} J.,    et al. 2001, \aj, 122, 1861

\bibitem[\protect\citeauthoryear{{Tabur}, {Kiss} \& {Bedding}}{{Tabur}
  et~al.}{2009}]{2009ApJ...703L..72T}
{Tabur} V.,  {Kiss} L.~L.,    {Bedding} T.~R.,  2009, \apjl, 703, L72

\bibitem[\protect\citeauthoryear{{van Gorkom}, {Knapp}, {Ekers}, {Ekers},
  {Laing} \& {Polk}}{{van Gorkom} et~al.}{1989}]{1989AJ.....97..708V}
{van Gorkom} J.~H.,  {Knapp} G.~R.,  {Ekers} R.~D.,  {Ekers} D.~D.,  {Laing}
  R.~A.,    {Polk} K.~S.,  1989, \aj, 97, 708

\bibitem[\protect\citeauthoryear{{van Gorkom}, {Knapp}, {Raimond}, {Faber} \&
  {Gallagher}}{{van Gorkom} et~al.}{1986}]{1986AJ.....91..791V}
{van Gorkom} J.~H.,  {Knapp} G.~R.,  {Raimond} E.,  {Faber} S.~M.,
  {Gallagher} J.~S.,  1986, \aj, 91, 791

\bibitem[\protect\citeauthoryear{{Vanzi} \& {Sauvage}}{{Vanzi} \&
  {Sauvage}}{2004}]{2004A&A...415..509V}
{Vanzi} L.,  {Sauvage} M.,  2004, \aap, 415, 509

\bibitem[\protect\citeauthoryear{{Williams}, {Dalcanton}, {Gilbert}, {Stilp},
  {Dolphin}, {Seth}, {Weisz} \& {Skillman}}{{Williams}
  et~al.}{2010}]{2010ApJ...716...71W}
{Williams} B.~F.,  {Dalcanton} J.~J.,  {Gilbert} K.~M.,  {Stilp} A.,  {Dolphin}
  A.,  {Seth} A.~C.,  {Weisz} D.,    {Skillman} E.,  2010, \apj, 716, 71

\bibitem[\protect\citeauthoryear{{Yi}, {Demarque} \& {Oemler}}{{Yi}
  et~al.}{1998}]{1998ApJ...492..480Y}
{Yi} S.,  {Demarque} P.,    {Oemler} A.~J.,  1998, \apj, 492, 480

\end{thebibliography}
\bibliographystyle{mn2e}

%%%%%%%%%%%%%%%%%%%%%%%%%%%%%%%%%%%%%%%%%%%%%%%%%%%%%%%%%%%%%%%%%%%%%%%

\bsp
\label{lastpage}
\end{document}